\documentclass[apj,twocolumn,twocolappendix,numberedappendix]{openjournal}

\usepackage{amsmath}
\usepackage{booktabs}
\usepackage{multirow}
\usepackage{color}
\usepackage{soul}
\usepackage{threeparttable}
\usepackage{float}
\usepackage{graphicx}
\usepackage{CJK}
\usepackage{xspace}
\usepackage{afterpage}
\usepackage{placeins}
\usepackage{amssymb}
\usepackage[breaklinks,colorlinks,citecolor=blue,urlcolor=blue,linkcolor=blue,filecolor=blue]{hyperref}
\usepackage[normalem]{ulem}
\usepackage{lineno}

\usepackage{placeins}

\makeatletter
\patchcmd{\NAT@citex}
  {\@citea\NAT@hyper@{%
    \NAT@nmfmt{\NAT@nm}%
    \hyper@natlinkbreak{\NAT@aysep\NAT@spacechar}{\@citeb\@extra@b@citeb}%
    \NAT@date}}
  {\@citea\NAT@nmfmt{\NAT@nm}%
  \NAT@aysep\NAT@spacechar\NAT@hyper@{\NAT@date}}{}{}

\patchcmd{\NAT@citex}
  {\@citea\NAT@hyper@{%
    \NAT@nmfmt{\NAT@nm}%
    \hyper@natlinkbreak{\NAT@spacechar\NAT@@open\if*#1*\else#1\NAT@spacechar\fi}%
      {\@citeb\@extra@b@citeb}%
    \NAT@date}}
  {\@citea\NAT@nmfmt{\NAT@nm}%
  \NAT@spacechar\NAT@@open\if*#1*\else#1\NAT@spacechar\fi\NAT@hyper@{\NAT@date}}
  {}{}
\makeatother

\shorttitle{Ghostly DLAs in SDSS DR16}
\shortauthors{Petitjean, P.}

\newcommand{\kms}{\ensuremath{\,\mathrm{km\,s^{-1}}}}
\newcommand{\orcidauthor}[3]{\author{\href{http://orcid.org/#1}{#2$^{#3}$}}}
\begin{document}

\title{\vspace{-1cm} Ghostly DLAs in SDSS DR16 \vspace{-1.75cm}}



\orcidauthor{0000-0000-0000-0000}{Patrick Petitjean}{1}

\affiliation{$^{1}$ Institut d'Astrophysique de Paris, CNRS, Sorbonne Universit\'e, 98bis Boulevard Arago, 75014 Paris, France}

\thanks{$^*$E-mail: \href{mailto:petitjean@iap.fr}{petitjean@iap.fr}}


\begin{abstract}
In Petitjean (2026), we revisited the origin of proximate damped Lyman-$\alpha$ absorbers (PDLAs), which trace cold gas within 3000~\kms\ of the quasar redshift, and interpreted their kinematics and physical properties within a unified framework. We showed that most PDLAs are associated with the environment of the AGN and/or the quasar host galaxy. We also provide the first census and characterization of absorption systems exhibiting strong absorption from excited levels of atomic ground states among quasar-associated absorbers.
Among these, ghostly and coronagraphic systems arise in dense, compact gas that partially covers the quasar emission regions. Most systems are associated with outflows reaching velocities up to $-2000$~\kms, while a smaller fraction of inflowing clouds extends to velocities of up to $+1200$~\kms.

In the present work, we provide an updated classification of PDLAs, including a revised catalogue of ghostly systems that more than doubles the number of previously known detections. We investigate the properties of these systems by measuring and discussing the equivalent widths of the detected absorption lines in both stacked spectra and individual ghostly systems. In particular, we show that although most ghostly systems are bona fide DLAs, this is not always the case.

\end{abstract}



\section{Introduction}

The feedback effect of AGN winds is thought to have a profound impact on galaxy formation by quenching or enhancing star-formation
\citep[e.g.][]{Springel2005,Cattaneo2009,Schaye2015,Dubois2016,Piotrowska2022,Bluck2023,Lauzikas2024}.

These winds are multi-scale and cover a wide range of ionization states and spatial scales \citep{Laha2021}. 
They are seen in X-rays as so-called warm absorbers \citep{Laha2014}
or  ultra-fast outflows (UFOs) \citep{Tombesi2011, Gofford2015}
with relativistic velocities of $\sim$0.1--0.3~c. 

In the UV and optical, 15 to 20\% of AGNs exhibit blue-shifted high ionization absorption lines (C~{\sc iv}, N~{\sc v}, O~{\sc vi}...)  often in the form of Broad Absorption Line (BAL) systems \citep{Weymann1981,Gibson2009}, with outflow velocities up to a few 10$^{3-4}$~\kms. The variations of the absorptions 
\citep{FilizAk2012, Aromal2024} argues for the gas being located close 
to the accretion disk although ionization arguments yield to much
larger distances \citep{Arav2020}.

Although a distinction between the outflows is made because
they are detected with different technics, they may share similar origin although not located at the same distance to the central AGN 
\citep[e.g.][]{Bu2021}.

AGN ionized winds are observed on much larger (galactic) scales \citep{Zakamska2016, Leung2019, ForsterSchreiber2019}.   
Same for neutral and molecular outflows thought to be the consequence of the action of AGN winds onto
the ISM of the host-galaxy.  
However how these large scale outflows that may quench star
formation in the galaxy are launched and how do the radiation and 
particle out-flows couple with the host galaxy gas is still a matter of debate \citep{Faucher2012}. 

Winds and radiation from the central AGN compress and stir the neutral gas in and around the host galaxy. It is therefore expected that some signature of this interaction should be imprinted in the quasar spectrum as H~{\sc i} absorption, in particular as proximate damped Lyman-alpha systems (DLAs) with small velocities relative to the quasar, typically $\Delta V < 3000$\kms \citep{Ellison2002, Prochaska2009} .
%
Previous studies concluded that most PDLAs arise
in gas associated with nearby galaxies, the gas
in the circumgalactic medium of the host galaxy being fully ionized
by the AGN radiation.
H~{\sc i} has however been observed in absorption in the halos of bright quasars within 100~kpc from the AGN \citep{Prochaska2009}.

In addition it is known that the clouds giving rise to proximate DLAs can sometimes
be sufficiently small that they do not fully cover the narrow-line
region \citep[so-called coronagraph DLAs or DLA-Cor, ][]{Hennawi2009,Finley2013,Jiang2016}, 
or even the broad-line region  \citep[so-called ghostly DLA,][]{Fathivavsari2017}. The latter clouds have been 
claimed to be located close to the AGN \citep{Laloux2021}. Finally, 
\citet{Noterdaeme2019} observed an excess of H$_2$ bearing PDLAs. 
They concluded that although most systems are likely associated with galaxies in the quasar environment, a small fraction may reside in the quasar host galaxy itself.

In \citet{Petitjean2026a}, we revisit the nature of high $N$(H~{\sc i})
proximate systems in light of recent observations and 
simulations, substantially increasing the available statistics thanks to the SDSS DR16 dataset. In the present research note, we 
provide additional observational information on the systems 
and give the list of ghostly DLAs.

\section{The different classes of metal ProxSys}

We call metal ProxSys, systems located 
within 3000~km~s$^{-1}$ from the quasar redshift and
detected by strong low-ionization metal absorptions.
The selection procedure of the metal ProxSys is described in Section 2 of 
\citet{Petitjean2026a}. Following is a quick summary.
  
If we wish to probe the full population 
of high $N$(H~{\sc i}) ProxSys, including all types of PDLAs
{\bf and in particular ghostly DLAs}, a special selection strategy 
is required, which does not use the presence of a H~{\sc i} Lyman-$\alpha$ trough.
Consequently, we select all absorption systems that can be identified through strong metal lines, without reference to their H~{\sc i} content.  

We use the final SDSS-IV quasar catalog from Data Release~16 (DR16) of the extended Baryon Oscillation Spectroscopic Survey \citep[eBOSS;][]{Dawson2016, Ahumada2020}, which we will refer to as DR16Q. This catalog contains 750,414 quasars \citep{Lyke2020}.

We impose that at least two low-ionization and two high-ionization lines
be detected at more than the 3~$\sigma$ level. The candidates are 
then visually inspected and classified. 

We classify the metal ProxSys into several categories, HighWSys, LowWSys, VeryLowWSys, DLA-cor, ghostly DLAs.
The first three categories, HighWSys, LowWSys and VeryLowWSys, correspond to systems with 
$W_{\rm r}$(Ly$\alpha$)$>$5~\AA, in the range 2-5~\AA~ and $<$2~\AA, respectively.
Given the medium spectral resolution of SDSS data, the low signal-to-noise ratio
and the fact that the Lyman-$\alpha$ trough is located in the wing of the 
Lyman-$\alpha$ emission from the quasar, this equivalent width is very uncertain
and although a number of HighWSys are bona fide DLAs, not all of them are.
Most of other systems are sub-DLAs. The DLA-Cor category comprises systems exhibiting strong residual features at the bottom of a clearly detected DLA trough, such as 
emission lines or significant continuum flux. Finally, ghostly DLAs correspond to systems in which the Ly-$\alpha$ absorption is absent or only weakly detected.

In addition to these categories, we singled out systems where absorption from
the excited level of the Si~{\sc ii} ground state is detected. 
In the following we will use four categories: HighWSys, HighWSys SiII*, 
DLA-Cor*, and ghostly DLAs. 
The HighWSys SiII* category corresponds to HighWSys with 
absorption from the excited level of the Si~{\sc ii} ground state detected. 
And same for DLA-Cor*. There are 1688, 54, 48 and 185 objects in the HighWSys,
HighWSys SiII*, DLA-Cor* and ghost DLA categories, respectively.
The systems with Si~{\sc ii}* detected represent about 15\%~of all 
systems.

We show in the following that the properties of these four 
classes of systems follow a clear trend with increasing 
strength of high ionization (O~{\sc vi}, N~{\sc v}, C~{\sc iv}) absorptions and increasing
strength of absorptions from excited atomic levels.

\section{Mean stacked spectra}
The full stacked spectra of the four categories we discuss here
are shown on Figs.~6 and 8 of \citet{Petitjean2026a}. Here we look at them in more details.

\begin{figure}[h]
   \centering
   \includegraphics[width=\linewidth, keepaspectratio]{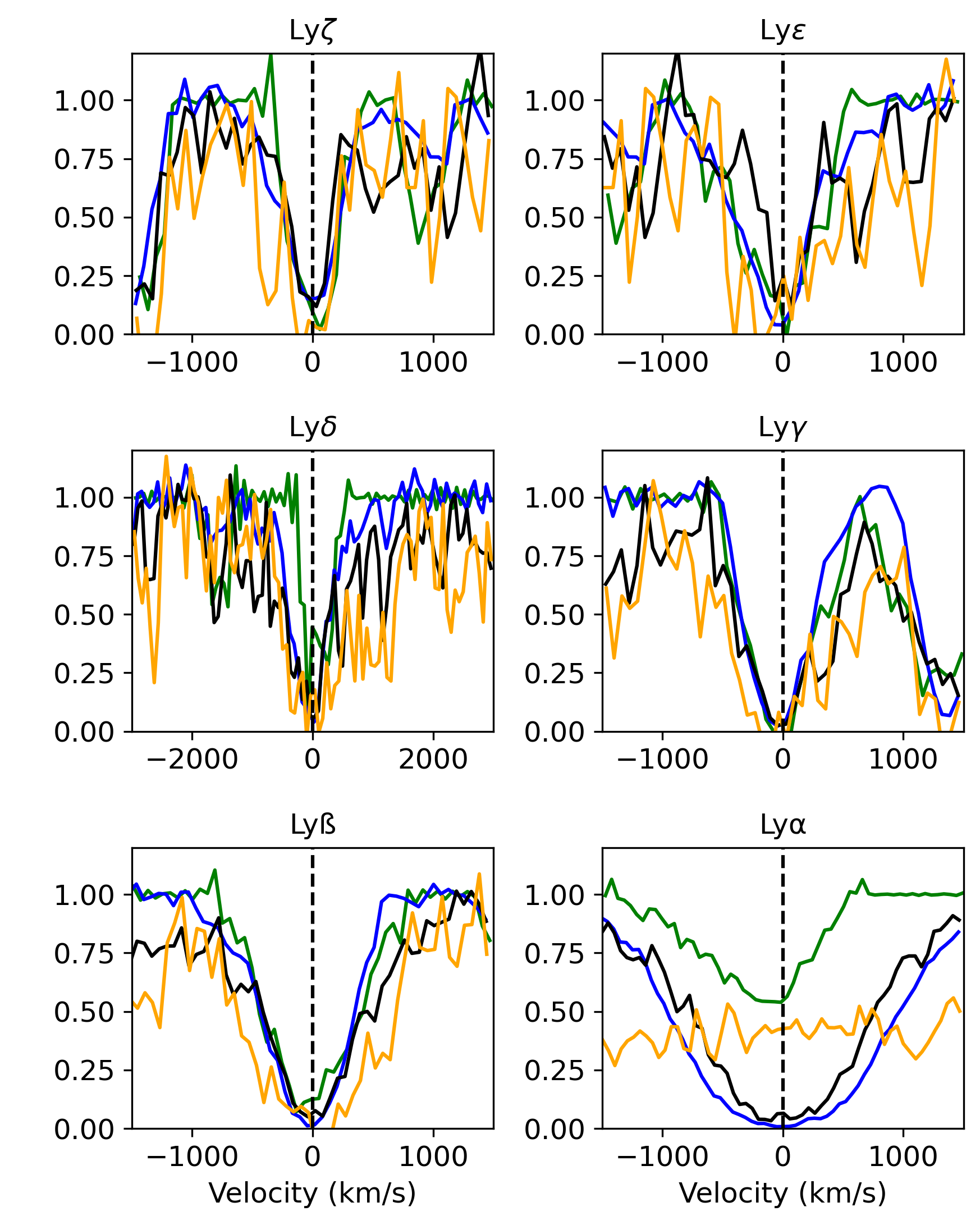}
      \caption{Mean spectra of the HighWSys (blue), HighWSys with Si II* (black), DLA-Cor with Si II* (orange), and ghostly DLA (green) systems, shown on a velocity scale for H~{\sc i} Lyman-$\alpha$, $\beta$, and $\gamma$. }
         \label{VitspeciesHI}
   \end{figure}

In Fig.~\ref{VitspeciesHI}, we present the H~{\sc i} absorption 
profiles for Lyman-$\alpha$, $\beta$, and $\gamma$ for the four 
composite spectra. It is 
apparent that the Lyman-$\alpha$ absorption is barely detectable in ghostly DLAs, when the corresponding Lyman-$\beta$ and Lyman-$\gamma$
absorptions are strong. In contrast, the damping wings are clearly visible in DLA-Cor systems, although a substantial residual flux remains at the bottom of the absorption trough. This residual is largely flat because the narrow emission arising from the uncovered portion of the narrow-line region is not always aligned with the absorption centroid. As a result, the stacking procedure smooths out the emission feature, producing a nearly uniform residual signal.

We also estimate the H~{\sc i} column density by fitting the Lyman series in the median spectra, excluding Lyman-$\alpha$ in the case of ghostly DLAs. Results are given in Table~~\ref{tab:EW}.
Relatively large Doppler parameters are required to reproduce the observed profiles of the higher-order Lyman-series lines. For small values of $b$, the intrinsically narrow saturated absorptions are significantly diluted by the SDSS spectral resolution, resulting in residual flux at line center. Larger Doppler parameters broaden the absorption cores and allow the convolved profiles to reach the observed zero-flux level. Note that large $b$-values reflect a significant velocity spread of the multi-components in the gas. The Doppler parameter is
larger for systems with detected absorption from excited
atomic levels (see also Table~\ref{tab:Vitspread}).

The inferred H~{\sc i} column densities for the HighWSys and HighWSys* classes are comparable and lower than those measured for the other absorber classes. DLA-Cor systems exhibit the highest H~{\sc i} column densities. Consistently, their H~{\sc i} Lyman-$\alpha$ equivalent widths are among the largest in the sample. This shows that these systems arise in dense, compact clouds. We further note that these results reinforce the conclusion that ghostly DLAs are generally bona fide DLAs (see, however, Section~\ref{sec: NHI}).

In Table~\ref{tab:EW}, we list the equivalent widths of the transitions detected in the stacked spectra. Caution should be exercised for transitions located in the Ly-$\alpha$ forest that are often blended. The associated uncertainties are estimated to be up to 0.3~\AA\ within the forest and 0.1~\AA\ outside it at the three sigma level.

We observe a clear trend from HighWSys to ghostly DLAs, whereby absorptions from high-ionization species (e.g., O~{\sc vi}, N~{\sc v}, C~{\sc iv}) and
transitions from excited levels of the ground state (e.g., Si~{\sc ii}* and 
C~{\sc ii}$^*$) become progressively stronger. We also note that Mg~{\sc ii}
absorption strengthens along this sequence, whereas Fe~{\sc ii} shows a much more modest increase.

   \begin{figure}[h]
   \centering
   \includegraphics[width=\linewidth, keepaspectratio]{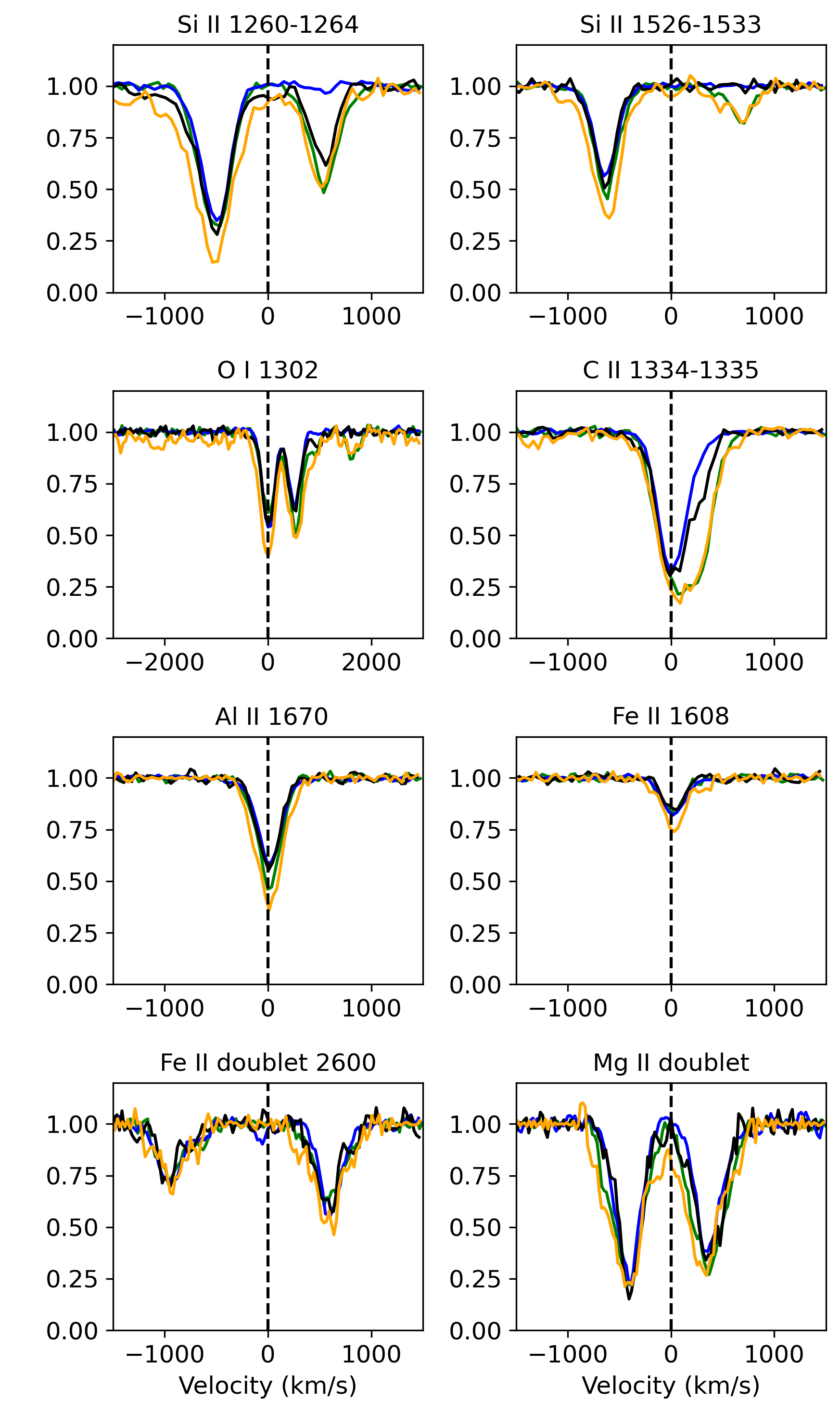}
      \caption{Mean spectra of the HighWSys (blue), HighWSys Si II* (black), DLA-Cor with Si II* (orange), and ghostly DLA (green) systems, shown on a velocity scale for low-ionization species. }
         \label{Vitspecieslowions}
   \end{figure}

\begin{figure}[h]
   \centering
   \includegraphics[width=\linewidth, keepaspectratio]{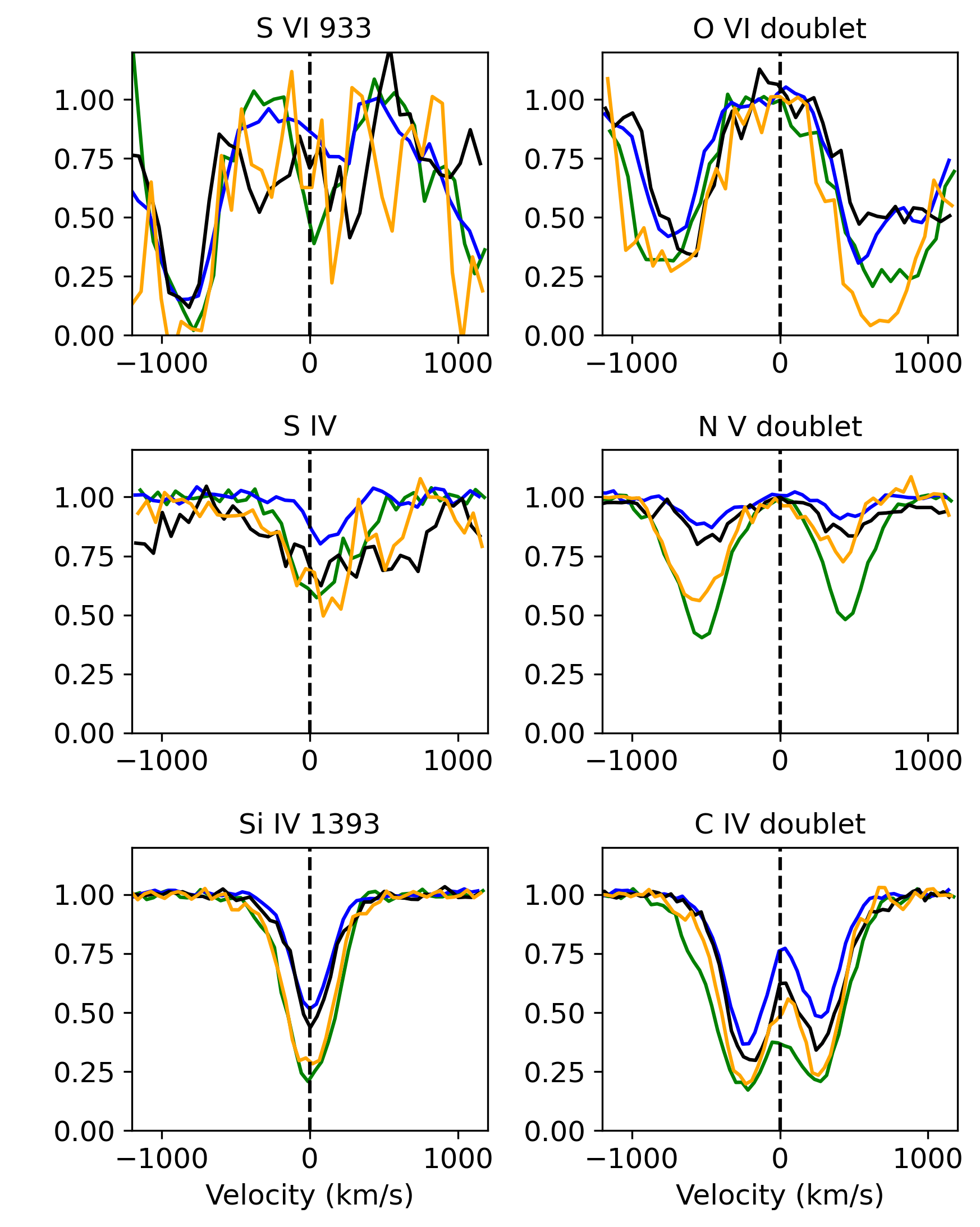}
      \caption{Mean spectra of the HighWSys (blue), HighWSys Si II* (black), DLA-Cor with Si II* (orange), and ghostly DLA (green) systems, shown on a velocity scale for high-ionization species }
         \label{Vitspecieshighions}
   \end{figure}

\begin{figure}[h]
   \centering
   \includegraphics[width=\linewidth, keepaspectratio]{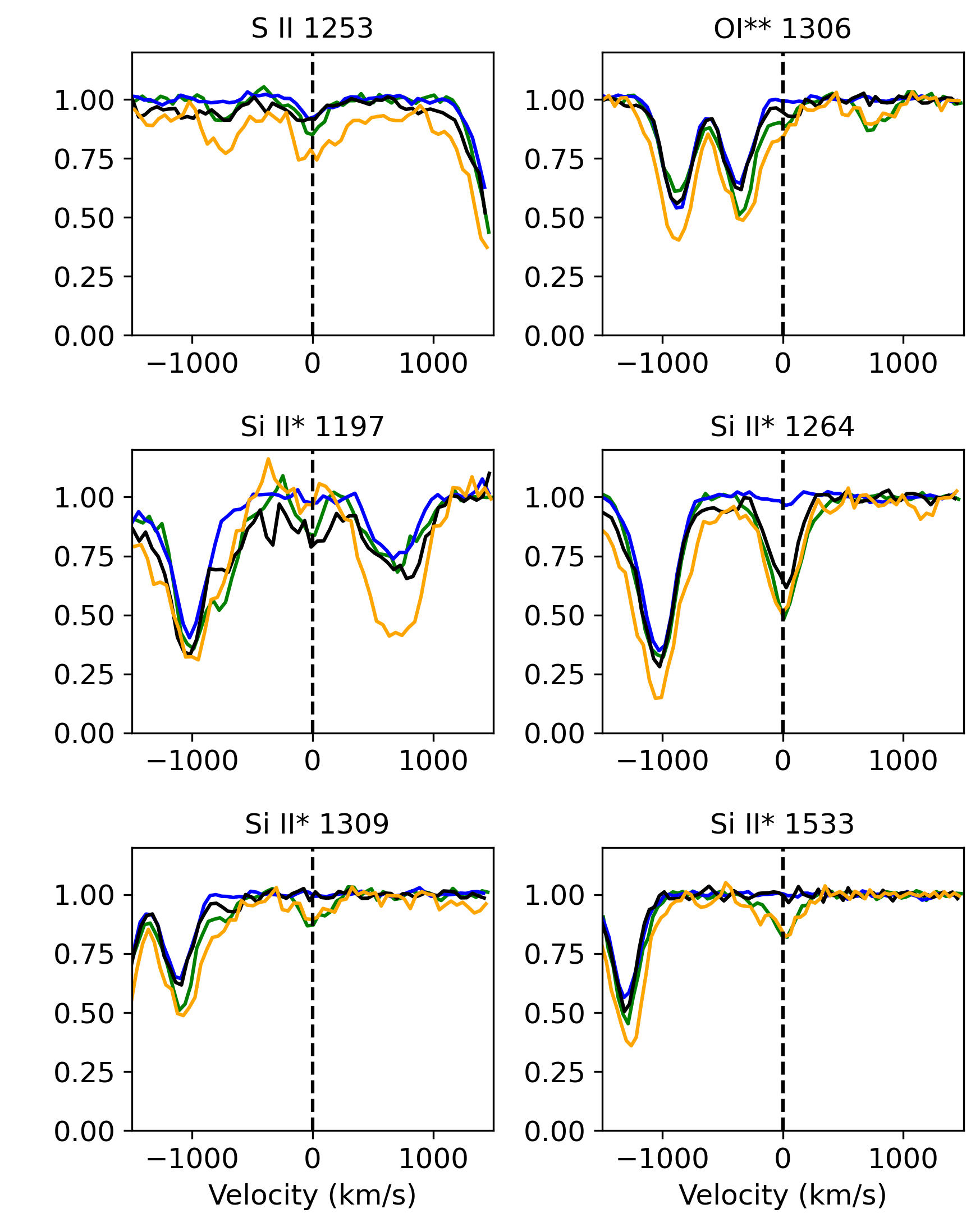}
      \caption{Mean spectra of the HighWSys (blue), HighWSys with Si II* (black), DLA-Cor with Si II* (orange), and ghostly DLA (green) systems, shown on a velocity scale for S~{\sc ii} and absorptions from excited atomic levels.}
         \label{VitspeciesExcited}
   \end{figure}

This is illustrated in Figs.~\ref{Vitspecieslowions}, \ref{Vitspecieshighions}, and \ref{VitspeciesExcited}, where we overlay, on a velocity scale, portions of the mean spectra for HighWSys (blue), HighWSys SiII* (black), DLA-Cor with SiII* (orange), and ghostly DLA (green) systems around several transitions of high-, low-, and excited-ionization species, respectively.

Absorption is generally strongest in the DLA-Cor SiII* systems, a natural consequence of their higher H~{\sc i} column densities. The trend described above is clearly seen in both high-ionization and excited transitions. We also note the presence of O~{\sc i}$^{**}$$\lambda$1306 in ghostly DLA and DLA-Cor* systems, which supports the conclusion of \citet{Fathivavsari2017} and \citet{Petitjean2026a} that these clouds must be dense \citep[see Fig.~6 of][]{Silva2002} with densities larger than 10$^{3}$~cm$^{-3}$.

\begin{figure}[h]
   \centering
   \includegraphics[width=\linewidth, keepaspectratio]{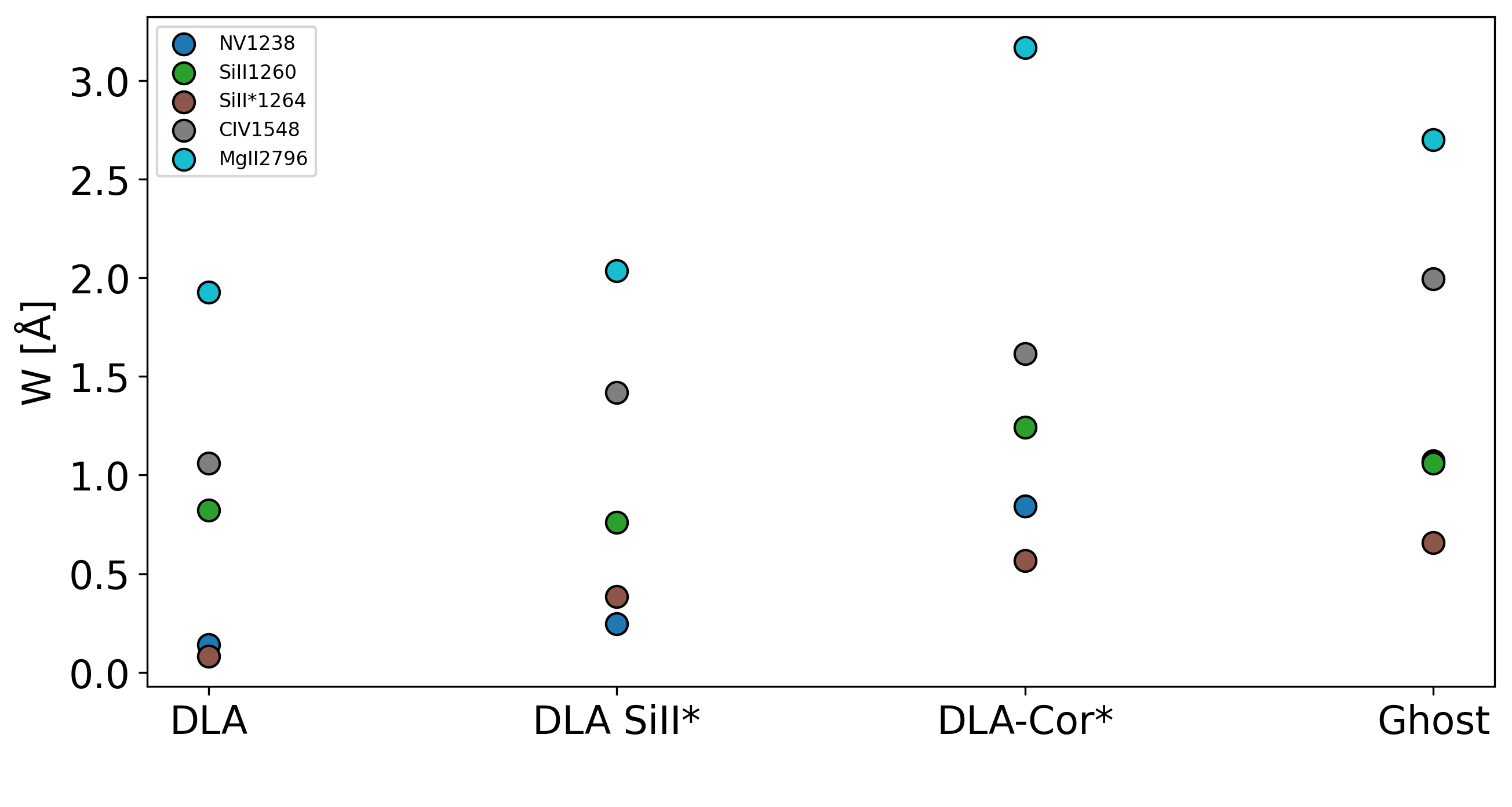}
      \caption{Equivalent width of transitions observed in the mean spectra of DLA, DLA with SiII*, DLA-Cor with SiII* and ghostly DLAs mean spectra.}
         \label{Wvsspectres}
   \end{figure}

This behaviour is further summarised in Fig.~\ref{Wvsspectres} where we plot the equivalent widths of a few important transitions for the four 
spectra considered here. We also list in Table~\ref{tab:Vitspread} the total velocity spreads of the Mg~{\sc ii}, C~{\sc iv}, and N~{\sc v} absorption lines for the four categories. This quantity is defined 
as the interval over which the optical depth is larger than 0.05.
The observed velocity spread of the gas increases markedly along the sequence. The corresponding values lie at the high end of those typically observed in intervening DLAs \citep[see e.g.][]{Zou2018}.


\begin{table}
\caption{Rest equivalent widths in \AA}
\label{tab:EW}
\begin{tabular}{lcccc}
\hline
 & DLA & DLA SiII* & DLA-Cor* & Ghost \\
\hline
log~NHI &  20.35 & 19.8 &   21.1 & 20.7\\    
b (km/s) &  105 & 126 & 137 &  127 \\   
SVI933 & 0.18: &   $<$0.1   & $<$0.1 & 0.82 \\
Ly-$\delta$ & 1.95 & 1.73 & 2.87 & 2.26 \\
Ly-$\gamma$ & 1.94 & 1.88 & 3.26 & 2.52 \\
CIII977 & 1.60 & 1.60 & 1.82 & 2.01 \\
OI988 & 0.79 & 0.55 & 0.86 & 0.61 \\
NIII989 & 0.53 & 0.75 & 1.36 & 1.06 \\
Ly-b & 2.70 & 2.73 & 4.07 & 3.02 \\
OVI1031 & 0.88 & 0.97 & 1.58 & 1.35 \\
CII-OVI-OI & 1.90 & 1.88 & 2.91 & 2.38 \\
SIV1062 & 0.19 & 0.49 & 0.64 & 0.75 \\
NII1083 & 0.48 & 0.59 & 1.53 & 1.01 \\
FeII1096 & 0.08 & 0.14 & 0.43 & 0.15 \\
FeIII1122 & 0.10 & 0.23 & 0.43 & 0.26 \\
NI1134 & 0.29 & 0.30 & 0.93 & 0.73 \\
SiII1190 & 0.64 & 0.65 & 1.12 & 1.23 \\
SiII1193 & 0.68 & 0.93 & 1.48 & 1.37 \\
NI1200 & 0.41 & 0.52 & 1.40 & 0.65 \\
SiIII1206 & 1.04 & 1.38 & 2.21 & 1.57 \\
Ly-a & 7.43 & 7.27 & 10.38 & 1.50 \\
NV1238 & 0.10 & 0.21 & 0.81 & 1.08 \\
NV1242 & 0.13 & 0.15 & 0.40 & 0.80 \\
SII1253 & 0.06 & 0.07 & 0.29 & 0.23 \\
SII1259 & 0.10 & 0.25 & 0.42 & 0.86 \\
SiII1260 & 0.80 & 0.75 & 1.20 & 1.09 \\
SiII*1264 & 0.04 & 0.43 & 0.58 & 0.66 \\
OI1302 & 0.56 & 0.47 & 0.87 & 0.58 \\
SiII1304 & 0.44 & 0.37 & 0.90 & 0.75 \\
NiII1317 & 0.02 & 0.01 & 0.07 & 0.11 \\
CII1334 & 0.98 & 1.33 & 2.00 & 2.10 \\
NiII1370 & $<$0.05 & $<$0.05 & 0.07 & 0.02 \\
SiIV1393 & 0.66 & 0.83 & 1.24 & 1.54 \\
SiIV1402 & 0.45 & 0.67 & 0.99 & 1.31 \\
SiII1526 & 0.52 & 0.63 & 1.12 & 0.91 \\
SiII*1533 & $<$0.05 & 0.04 & 0.30 & 0.27 \\
CIV1548 & 1.08 & 1.39 & 1.66 & 1.99 \\
CIV1550 & 0.83 & 1.27 & 1.54 & 1.98 \\
FeII1608 & 0.24 & 0.20 & 0.27 & 0.22 \\
CI1656 & $<$0.05 & $<$0.05 & $<$0.05 & 0.08 \\
AlII1670 & 0.56 & 0.63 & 1.15 & 0.89 \\
AlIII1854 & 0.20 & 0.27 & 0.65 & 0.76 \\
AlIII1862 & 0.13 & 0.15 & 0.46 & 0.48 \\
ZnII2026 & $<$0.05 & $<$0.05 & 0.20 & 0.11 \\
FeII2344 & 0.57 & 0.44 & 1.00 & 0.75 \\
FeII2374 & 0.29 & 0.19 & 0.34 & 0.29 \\
FeII2382 & 0.83 & 0.83 & 1.34 & 0.97 \\
Mn2576 & $<$0.05 & $<$0.05 & $<$0.05 & $<$0.05 \\
FeII2586 & 0.64 & 0.47 & 0.59 & 0.67 \\
MnII2594 & 0.11 & $<$0.05 & $<$0.05 & 0.09 \\
FeII2600 & 0.81 & 0.85 & 1.35 & 1.09 \\
MgII2796 & 1.89 & 2.04 & 3.17 & 2.68 \\
MgII2803 & 1.46 & 1.84 & 2.67 & 2.56 \\
MgI2852 & 0.38 & 0.36 & 0.75 & 0.75 \\
\hline
\end{tabular}
\end{table}
\begin{table}
\caption{Velocity spread in km/s}
\label{tab:Vitspread}
\begin{tabular}{lcccc}
\hline
 & HighWSys & HighWSys SiII* & DLA-Cor SiII* & Ghost \\
\hline
Mg II   &     510      &      590    &      786     &       672 \\
C IV    &     701      &      677    &     1259     &      1050  \\
N V     &     400      &      484    &      660     &       852   \\
\hline
\end{tabular}
\end{table}

\section{Ghostly DLAs}
\subsection{List of Ghostly DLAs}
During the initial visual inspection of all ProxSys, we classified systems showing no or little H~{\sc i} 
Lyman-$\alpha$ absorption as ghostly DLAs (Ghost) or probable ghostly DLAs (Ghost?) when some ambiguity was introduced by low signal-to-noise ratio and/or blending.

Among the 220 systems initially classified as Ghost (90) or Ghost? (130), we confirmed 95 as Ghost systems and 92 as probable Ghosts, while rejecting 33 after a more detailed inspection based on all available absorption features. We then stacked the Ghost and Ghost? spectra and found that they exhibit very similar properties. In particular, little to no absorption is detected at the expected position of 
the H~{\sc i} Lyman-$\alpha$ line, while strong absorption 
from excited atomic levels is clearly present. We therefore retained the full sample of 187 systems for further analysis.

Table~\ref{tab:ListGhost} presents a subset of the Ghost catalog containing 
the 187 objects. 
The columns are as follows: 
(1) \texttt{Name}: SDSS object identifier; 
(2–4) \texttt{MJD, Plate, Fiber}:  Modified Julian Date, spectroscopic plate ID, and fiber number; 
(5) \texttt{zQSO}: quasar redshift; 
(6) \texttt{g\_mag}: quasar magnitude in the g-band; 
(7) \texttt{Velocity}: system velocity relative to quasar; 
(8) \texttt{Flag}: object classification.
The classification flag is defined as follows: objects labeled \texttt{y} are considered secure detections, 
\texttt{y?} indicates most probable detections, \texttt{?} corresponds to probable cases.

We checked whether we recover the ghostly DLAs detected by \citet{Fathivavsari2020}. Out of the 89 systems, we recover 58. Of the 31 missing systems, seven (7) are misidentified ghostly DLAs, nine (9) are questionable, and nine (9) lie outside our search range (four with AI$>$1000~km/s and five with $z$$<$2.11). Therefore, we miss six confirmed ghostly DLAs, yielding a success rate of 86\%. Among these six, three have no C~{\sc ii} line detected but are confirmed by strong Mg~{\sc ii} absorption, while three show no high-ionisation absorption lines. This explains why these systems were missed. On the other hand, we find 36 additional confirmed ghostly DLAs and 94 probable ones. We 
therefore more than double the number of these systems.

The complete quasar catalogue is available in electronic form at the CDS and will be accessible through the VizieR catalogue service.

\FloatBarrier
\begin{table*}[htbp]
\centering
\caption{Excerpt from the Ghost catalog. Only the first 10 rows are displayed. The full list is available on VizieR. }
\begin{tabular}{l c c c c c c c}
\hline
Name & MJD & Plate & Fiber & zQSO & g\_mag & Vitesse & Flag \\
(1)  & (2) & (3)   &  (4)  &  (5) &   (6)  &   (7)   &   (8)   \\
\hline
000125.69+141022.3 & 56268 & 6177 & 318 & 2.421 & 20.19 & -2860 & y \\
000818.48+043249.7 & 57663 & 8743 & 924 & 3.377 & 19.86 & -2960 & ? \\
000943.73+132032.6 & 56191 & 6112 & 414 & 2.390 & 21.02 & -1250 & y \\
000958.65+015755.1 & 55511 & 4298 & 392 & 2.973 & 19.32 &   140 & y \\
001245.12-054945.7 & 56569 & 7036 & 370 & 2.214 & 20.41 & -1150 & y \\
001316.82-093841.2 & 56628 & 7169 & 280 & 2.644 & 19.37 & -2310 & y \\
002039.59-030722.6 & 57358 & 7900 & 958 & 2.220 & 20.71 & -1550 & y \\
003606.10+272539.2 & 57360 & 7671 & 433 & 2.180 & 21.28 & -1970 & ? \\
003901.47+073434.2 & 55882 & 4541 & 860 & 2.267 & 21.15 &   590 & ? \\
003914.11+312920.9 & 57364 & 7715 & 193 & 2.200 & 20.04 &   140 & y \\ 
\hline
\end{tabular}
\label{tab:ListGhost}
\end{table*}
 
\subsection{H~{\sc i} column densities in Ghostly DLA}
\label{sec: NHI}
It is difficult to measure the H~{\sc i} column density in individual ghost systems for several reasons: (i) the Lyman-$\alpha$ line is not available; (ii) depending on redshift, only a limited number of higher-order Lyman-series transitions can be used; and (iii) the signal-to-noise ratio is typically low in the wavelength range where lines beyond Lyman-$\beta$ are redshifted.

\begin{figure}[h]
   \centering
   \includegraphics[width=\linewidth, keepaspectratio]{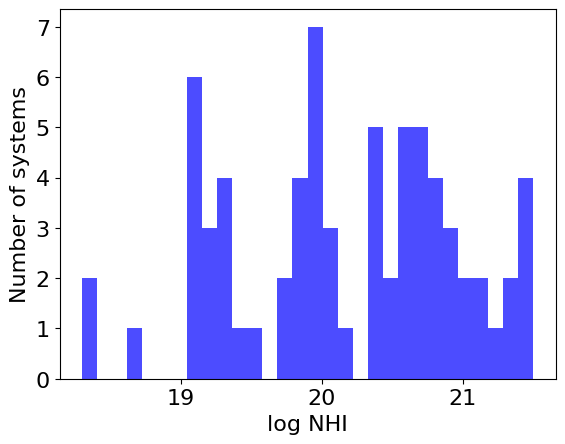}
      \caption{Distribution of log$N$(H~{\sc i}) measured in individual ghostly systems from fitting the Lyman series, excluding Lyman-$\alpha$.}
         \label{HistoLogNHIGhost}
   \end{figure}

Nevertheless, we have estimated column densities whenever possible. These values should be regarded as indicative only. Although the fits are somewhat degenerate, the inferred Doppler parameters are often larger than 50~km~s$^{-1}$. The results of the fits are shown in Fig.~\ref{HistoLogNHIGhost}. Interestingly, about one third of the systems have column densities robustly below 10$^{20}$~cm$^{-2}$. The relatively high column density inferred from the median spectrum may be overestimated due to the smearing of absorption features associated with large Doppler parameters.

%
\subsection{Absorption equivalent widths}
\label{sec: Correlations}
\begin{figure}[h]
   \centering
   \includegraphics[width=\linewidth, keepaspectratio]{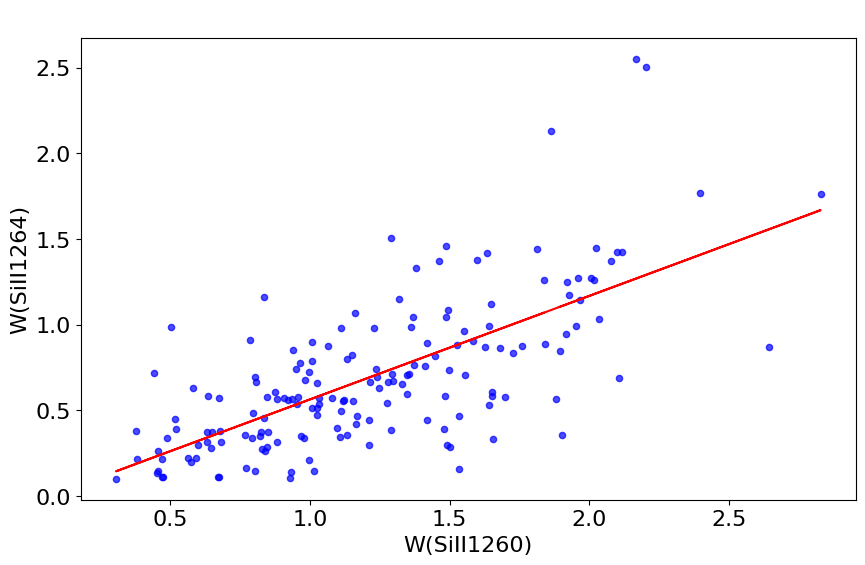}
      \caption{The equivalent width of Si~{\sc ii}*$\lambda$1264 versus 
      the equivalent width of Si~{\sc ii}$\lambda$1260 in individual 
      ghostly DLAs.}
         \label{SiII}
   \end{figure}

In Fig.~\ref{SiII}, we show the equivalent width of Si~{\sc ii}$^*$$\lambda$1264 as a function of the equivalent width of Si~{\sc ii}$\lambda$1260. The two quantities are strongly correlated with a Pearson coefficient of 0.69 and a p-probability smaller than
10$^{-10}$. This is likely a consequence of significant saturation in both transitions at least for part of 
the absorption profile. In this regime, variations in equivalent width primarily reflect differences in the velocity structure rather than in column density. This also suggests 
that Si~{\sc ii} excitation is present in most but not all
components of the absorbing complexes.

\begin{figure}[h]
   \centering
   \includegraphics[width=0.85\linewidth, keepaspectratio]{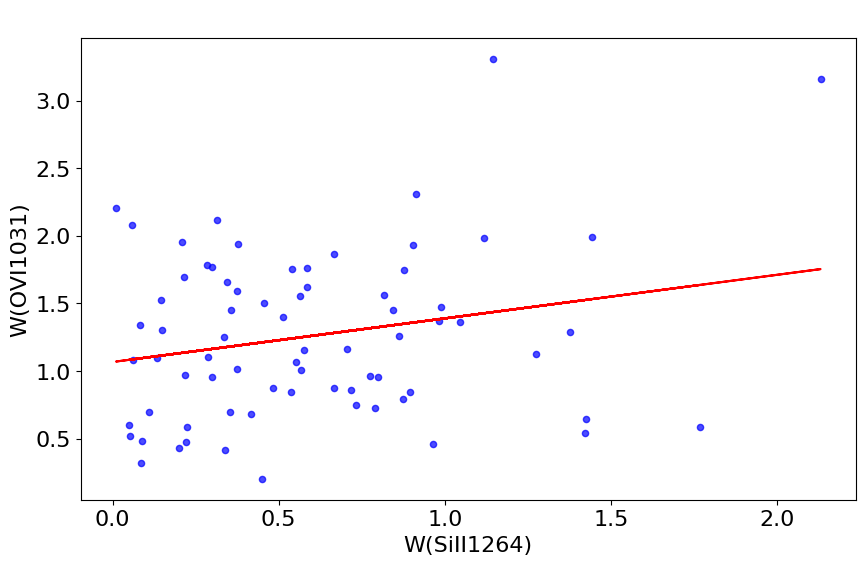}
   \includegraphics[width=0.85\linewidth, keepaspectratio]{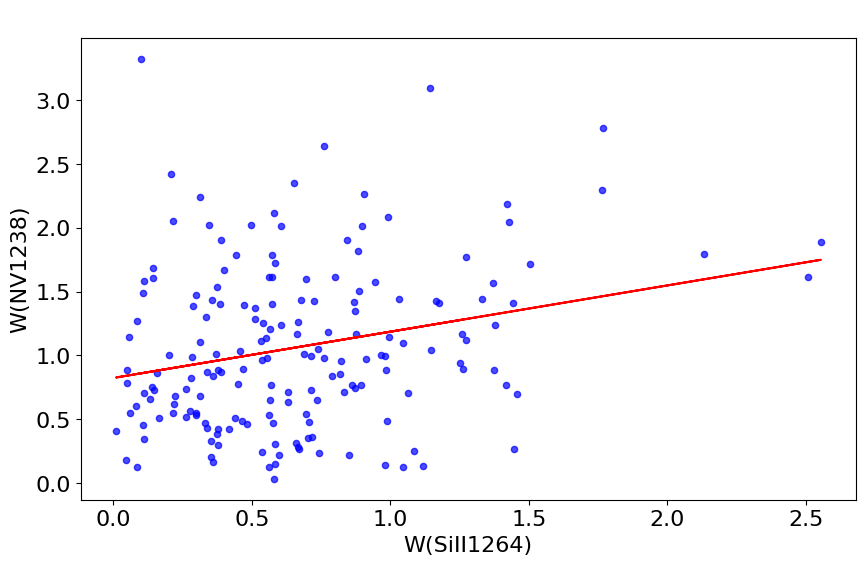}
   \includegraphics[width=0.85\linewidth, keepaspectratio]{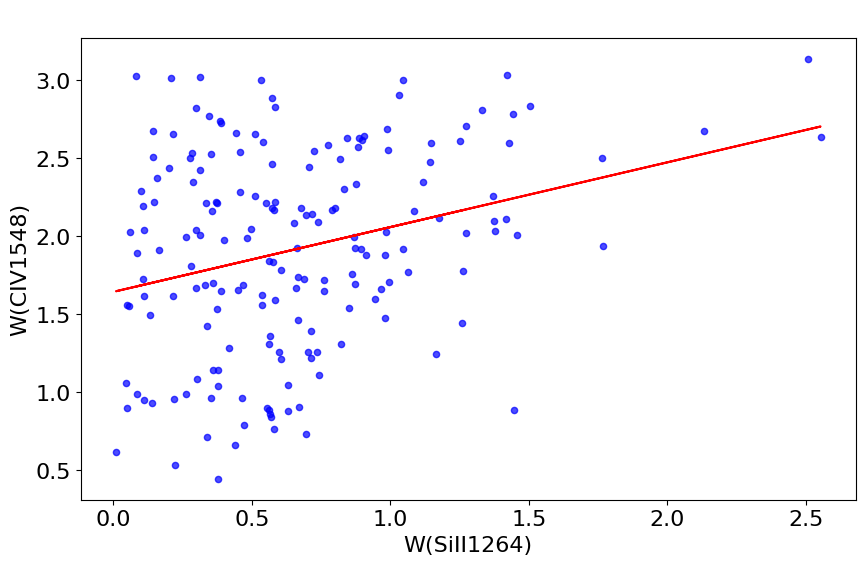}   
      \caption{The equivalent widths of C~{\sc iv}$\lambda$1548 (bottom), N~{\sc v}$\lambda$1238 (middle) and O~{\sc vi}$\lambda$1031 (top)  as a function of the equivalent width of Si~{\sc ii}$*\lambda$1264. }
         \label{Highexcitation}
   \end{figure}

We investigate the relation between the excitation of Si~{\sc ii} and the presence of high-ionization species (C~{\sc iv}, N~{\sc v}, and O~{\sc vi}). In Fig.~\ref{Highexcitation}, we show in different panels the equivalent widths of these three high-ionization species as a function of the equivalent width of Si~{\sc ii}*$\lambda$1264. A mild but statistically significant correlation is observed for all three ions, with Pearson coefficients of 0.29, 0.25, and 0.23, and corresponding p-values of $8.8 \times 10^{-5}$, $6.3 \times 10^{-4}$, and 0.05, respectively. 

The presence of strong high-ionization absorption indicates that the gas is located in the vicinity of the quasar \citep[see also][]{Fathivavsari2017}. The coexistence of low-ionization species requires relatively high gas densities, a conclusion further supported by the strong Si~{\sc ii}* absorption, whose excitation is expected to be dominated by collisional processes. 
{\bf Although it is difficult to 
estimate the exact density, it is expected to be of order  a few 100--1000~cm$^{-3}$} \citep[see,][]{Silva2002}.

\subsection{Kinematics}
\begin{figure}[h]
   \centering
   \includegraphics[width=0.85\linewidth, keepaspectratio]{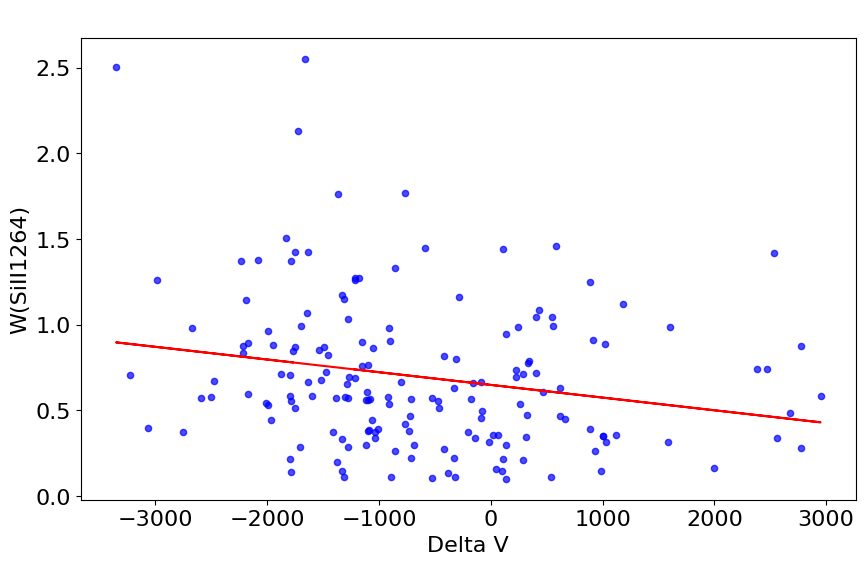}
      \caption{The equivalent width of Si~{\sc ii}*$\lambda$1264 
as a function of the velocity of the system relative to the quasar 
redshift (in km~s$^{-1}$).}
         \label{SiII*vsV}
   \end{figure}

In Fig.~\ref{SiII*vsV}, we show the equivalent width of the
Si~{\sc ii}$*\lambda$1264 transition as a function of the velocity 
of the system relative to the quasar redshift. A trend is apparent, whereby the level of excitation increases with increasing velocity offset, consistent with outflowing gas. The Pearson correlation coefficient is $-0.21$, with a corresponding p-value of 0.005.

We have shown in \citet{Petitjean2026a} that the kinematics of ghostly systems is characterized by two main components of approximately equal weight: one centered on
the quasar redshift, and another associated with an outflow spanning velocities
between $-1000$ and $-2000$~km~s$^{-1}$. The above anti-correlation thus supports our 
previous findings that ghostly DLAs trace outflowing gas resulting from the 
interaction between the interstellar medium of the host galaxy and the quasar-driven 
wind and radiation pressure.

\section{Conclusions}

In this research note, we have presented additional observational constraints on the nature of metal-selected proximate absorption systems identified in SDSS DR16 and provided a revised catalogue of ghostly DLAs. Our updated sample contains 187 ghostly DLA candidates, more than doubling the number of systems previously known.

By comparing the stacked spectra of HighWSys, HighWSys SiII*, DLA-Cor*, and ghostly DLA systems, we find evidence for a continuous sequence characterized by increasing strengths of high-ionization species (O~{\sc vi}, N~{\sc v}, and C~{\sc iv}), stronger absorption from excited atomic levels, and larger velocity spreads. The velocity spreads observed in DLA-Cor* and ghostly DLAs are among the largest known for H~{\sc i}-rich absorbers.

The detection of excited transitions such as Si~{\sc ii}* and O~{\sc i}$^{**}$, together with the observed correlations between excitation and high-ionization absorption strengths, indicates that the gas is both dense and exposed to an intense ionizing radiation field. These results support a picture in which the absorbers arise in compact clouds located in the 
vicinity of the AGN. The correlation between Si~{\sc ii}* excitation and velocity offset provides additional evidence that the clouds are 
preferentially associated with an outflowing component.

Finally, although the stacked spectra indicate H~{\sc i} column densities typical of the DLA regime, measurements performed on individual ghostly systems suggest a more complex picture. While most ghostly absorbers are bona fide DLAs, roughly one third of the systems appear to have column densities below the classical DLA threshold. Ghostly DLAs therefore represent a heterogeneous population of dense, compact absorbers located close to quasars and provide a unique probe of the interaction between AGN-driven winds and the surrounding interstellar medium.

\begin{acknowledgements}
This work was supported in part by the Agence Nationale de la Recherche (ANR, France) under contract ANR-22-CE31-0009.
We thank Camille No\^us (Laboratoire Cogitamus) for inappreciable and often unnoticed discussions, advice and support.
Funding for the Sloan Digital Sky Survey IV has been provided by the Alfred P. Sloan Foundation, the U.S. Department of Energy Office of Science, and the Participating Institutions. 

\end{acknowledgements}

%
\bibliographystyle{bibtex/aa}
\bibliography{macros/Master}

\end{document}